\documentclass[%
 reprint,
superscriptaddress,
 amsmath,amssymb,
 aps,
prd,
longbibliography
]{revtex4-1}

\usepackage{graphicx}
\usepackage{dcolumn}
\usepackage{bm}
\usepackage{hyperref}
\usepackage{color}
\usepackage[mathlines]{lineno}
\usepackage{tabularx}
\usepackage{subfigure}
\usepackage[normalem]{ulem}
\usepackage[export]{adjustbox}
\usepackage{url}
\hypersetup{
    pdfnewwindow=true,    
    colorlinks=true,      
    linkcolor=blue,       
    citecolor=blue,       
    filecolor=blue,       
    urlcolor=blue         
}

\newcommand\chandra{{Chandra}}
\newcommand\xmm{{XMM-Newton}}
\newcommand\suzaku{{Suzaku}}
\newcommand\integral{{INTEGRAL}}

\newcommand\nustar{\hbox{NuSTAR}}
\newcommand\fermi{{Fermi}}

\definecolor{forestgreen}{rgb}{0.13, 0.55, 0.13}
\newcolumntype{Y}{>{\centering\arraybackslash}X}
\usepackage{array}
\newcolumntype{C}{>{\centering\arraybackslash}p{3.8cm}}
\newcolumntype{D}{>{\centering\arraybackslash}p{6cm}}

\usepackage{natbib}

\begin{document}
\title{NuSTAR tests of sterile-neutrino dark matter: \\ New Galactic bulge observations and combined impact}

\author{Brandon M. Roach
}
\email{roachb@mit.edu}
\thanks{\href{orcid.org/0000-0001-8016-2170}{orcid.org/0000-0001-8016-2170}}
\affiliation{Department of Physics, Massachusetts Institute of Technology, Cambridge, MA 02139, USA}
\author{Kenny C. Y. Ng
}
\email{chun-yu.ng@weizmann.ac.il}
\thanks{\href{orcid.org/0000-0001-8016-2170}{orcid.org/0000-0001-8016-2170}}
\affiliation{{Department of Particle Physics and Astrophysics, Weizmann Institute of Science, Rehovot, Israel}}
\author{Kerstin Perez
}
\email{kmperez@mit.edu}
\thanks{\href{orcid.org/0000-0002-6404-4737}{orcid.org/0000-0002-6404-4737}}
\affiliation{Department of Physics, Massachusetts Institute of Technology, Cambridge, MA 02139, USA}
\author{John~F.~Beacom
}
\email{beacom.7@osu.edu}
\thanks{\href{orcid.org/0000-0002-0005-2631}{orcid.org/0000-0002-0005-2631}}
\affiliation{Center for Cosmology and AstroParticle Physics (CCAPP), Ohio State University, Columbus, OH 43210, USA}
\affiliation{Department of Physics, Ohio State University, Columbus, OH 43210, USA}
\affiliation{Department of Astronomy, Ohio State University, Columbus, OH 43210, USA} 
\author{Shunsaku Horiuchi
}
\email{horiuchi@vt.edu}
\thanks{\href{orcid.org/0000-0001-6142-6556}{orcid.org/0000-0001-6142-6556}}
\affiliation{Center for Neutrino Physics, Department of Physics, Virginia Tech, Blacksburg, VA 24061, USA}
\author{Roman Krivonos
}
\email{krivonos@iki.rssi.ru}
\thanks{\href{orcid.org/0000-0003-2737-5673}{orcid.org/0000-0003-2737-5673}}
\affiliation{Space Research Institute of the Russian Academy of Sciences (IKI), Moscow 117997, Russia}
\author{Daniel R. Wik
}
\email{wik@astro.utah.edu}
\thanks{\href{orcid.org/0000-0001-9110-2245}{orcid.org/0000-0001-9110-2245}}
\affiliation{Department of Physics and Astronomy, University of Utah, Salt Lake City, UT 84112, USA}

\date{Received 4 October 2019; accepted 2 April 2020; published 8 May 2020}

\begin{abstract}
We analyze two dedicated \nustar{} observations with exposure ${\sim}190\text{\,ks}$ located ${\sim}10^\circ$ from the Galactic plane, one above and the other below, to search for x-ray lines from the radiative decay of sterile-neutrino dark matter. These fields were chosen to minimize astrophysical x-ray backgrounds while remaining near the densest region of the dark matter halo. We find no evidence of anomalous x-ray lines in the energy range 5--20\,keV, corresponding to sterile neutrino masses 10--40\,keV. Interpreted in the context of sterile neutrinos produced via neutrino mixing, these observations provide the leading constraints in the mass range 10--12\,keV, improving upon previous constraints in this range by a factor ${\sim}2$. We also compare our results to Monte Carlo simulations, showing that the fluctuations in our derived limit are not dominated by systematic effects. An updated model of the instrumental background, which is currently under development, will improve \nustar{}'s sensitivity to anomalous x-ray lines, particularly for energies 3--5\,keV.
\newline
\newline
\textsc{doi}: \scriptsize{\href{https://doi.org/10.1103/PhysRevD.101.103011}{https://doi.org/10.1103/PhysRevD.101.103011}} 
\end{abstract}

\maketitle

\section{\label{sec:intro} Introduction}
\par Multiple lines of cosmological evidence indicate that ${\sim}80\%$ of the matter density of the Universe, and ${\sim}25\%$ of its energy density, is nonbaryonic and nonluminous, hence its name, dark matter (DM) \cite{Tanabashi:18}. At present, the effects of DM are only measurable via its gravitational effects on astronomical scales, ranging from the motions of galaxies and galaxy clusters to the power spectrum of the Cosmic Microwave Background \cite{Bertone:2004pz,Strigari:2013iaa,Seigar:2015,Buckley:2017ijx,Aghanim:2018eyx,Wechsler:2018}. The lack of a viable Standard Model candidate for particle DM (hereafter symbolized $\chi$) has led to a plethora of theoretical models, many of which are also motivated by a desire to account for other phenomena not explained by the Standard Model (e.g., baryogenesis, neutrino masses, the hierarchy problem, etc).

\par The techniques of indirect detection use astronomical observations to search for the decay and/or annihilation of DM into Standard Model particles such as electrons/positrons, (anti)protons/nuclei, neutrinos, and photons \cite{Gaskins:16}. Because photons are not deflected by astrophysical magnetic fields, it is possible to determine their arrival direction within the angular resolution of the detector, allowing for a rejection of photons from known astrophysical sources. Final states with mono-energetic photons are particularly valuable for indirect DM searches, as they result in line-like signals atop a (usually) smooth continuum background.

\par A popular DM candidate with $m_\chi \sim \text{keV}$ is the sterile neutrino, with models such as the $\nu\text{MSM}$ providing explanations for the particle nature of DM, neutrino masses, and baryogenesis \cite{Asaka:2005an,Asaka:2006nq,Canetti:2012kh,Canetti:2012vf}. The radiative decay of sterile neutrinos via $\chi \rightarrow \nu + \gamma$ would produce a mono-energetic x-ray photon and an active neutrino, each with $E = m_\chi\,/\,2$ \cite{Kusenko:2009up,Adhikari:2016bei,Abazajian:2017tcc,Boyarsky:2018tvu,Shrock:1974nd,Pal:1981rm,Dolgov:2000ew,Abazajian:2001vt}. 
\par
\begin{figure}[h!]
    \centering
    \includegraphics[width=0.98\columnwidth]{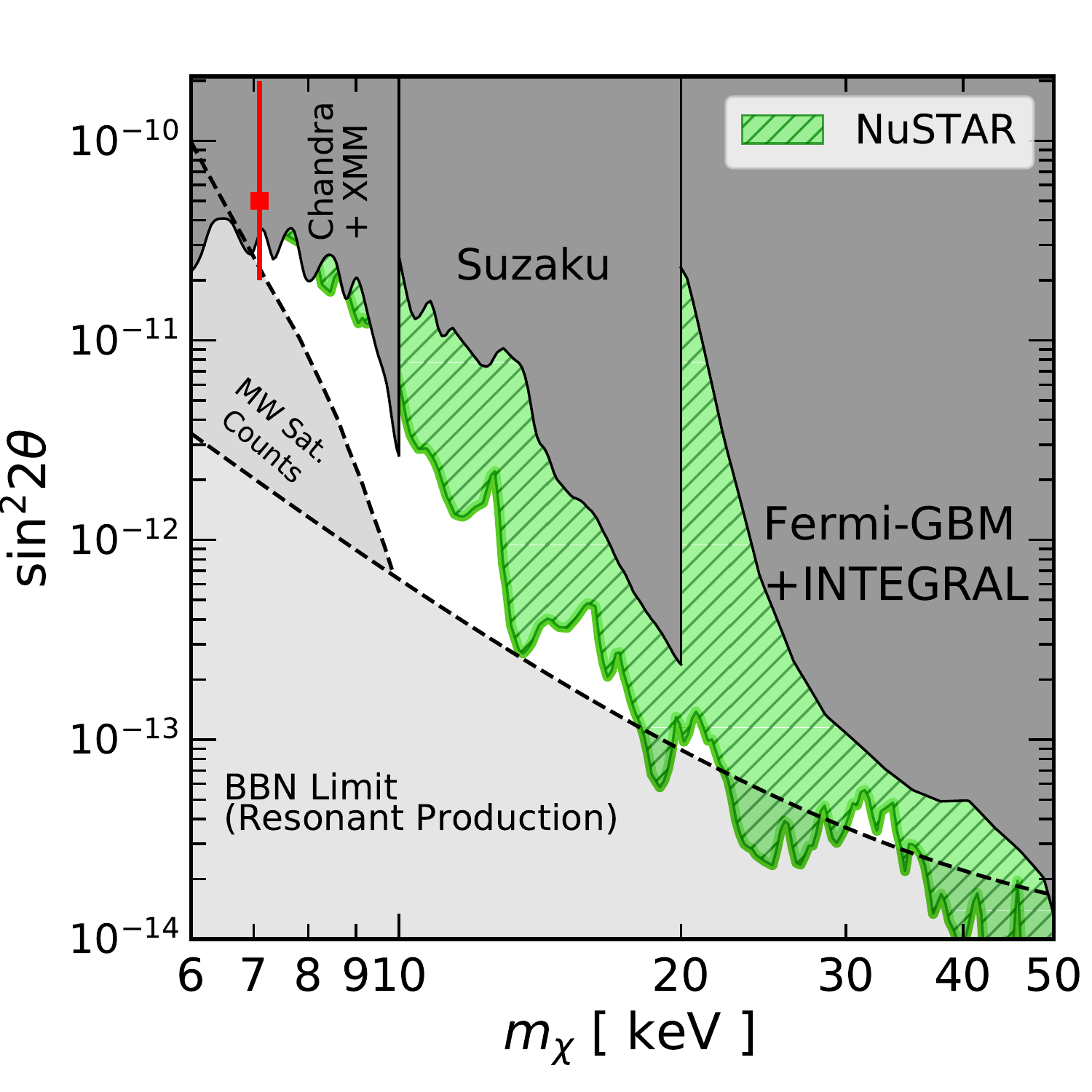}
    \caption{The combined impact on the $\nu\text{MSM}$ parameter space of previous \nustar{} searches \cite{Riemer-Sorensen:2015kqa,Perez:2016tcq,Neronov:2016wdd,Ng:2019a} and this work is indicated by the green region. This work provides the leading constraints in the 10--12\,keV mass range, as shown in Fig. ~\ref{fig:dmgamma_both_figs}. The tentative $E\simeq 3.5\text{\,keV}$ signal \cite{Boyarsky:2014jta,Bulbul:2014sua,Boyarsky:2014ska} is indicated by the red point. Constraints from other x-ray instruments \cite{Horiuchi:2015qri,Ruchayskiy:2015onc,Tamura:2014mta,Ng:2015gfa,Boyarsky:2007ge} are shown for comparison. Uncertainties associated with MW satellite counts~\cite{Cherry:2017dwu} and BBN~\cite{Venumadhav:2015pla,Laine:2008pg} are discussed in Sec.~\ref{sec:intro}.} 
    \label{fig:SNDM_NuSTAR_best}
\end{figure}

 Sterile neutrinos may be produced in the early Universe via mixing with active neutrinos \cite{Dodelson:1993je}, and this production may be resonantly enhanced by primordial lepton asymmetry \cite{Shi:1998km}. Considerations from big bang nucleosynthesis (BBN) \cite{Dolgov:2002ab, Serpico:2005bc, Boyarsky:2009ix} provide an upper bound on the cosmological lepton asymmetry per unit entropy density $L_6 \equiv 10^6(n_\nu - n_{\bar{\nu}})/s \leq 2500$, which we translate into the constraint on the active-sterile mixing angle $\sin^2 2\theta$ shown in Fig.~\ref{fig:SNDM_NuSTAR_best} using the \textsc{sterile-dm} code \cite{Venumadhav:2015pla}. We note that these BBN limits are particularly sensitive to the treatment of neutrino opacities and the plasma equation of state near the QCD phase transition, with different calculations finding different results---for example, the limits shown in Refs.~\cite{Laine:2008pg,Boyarsky:2009ix} for the same value of $L_6$ are nearly an order of magnitude less constraining than those from Ref.~\cite{Venumadhav:2015pla}, which the authors of Refs.~\cite{Cherry:2017dwu,Boyarsky:2018tvu} attribute to differences in the treatment of neutrino opacities in the QCD epoch. (An update to the calculation 
in Ref.~\cite{Laine:2008pg} is presented in Ref.~\cite{Ghiglieri:2015jua}, though the latter does not present an updated constraint in the $m_\chi-\sin^2 2\theta$ plane.) This lower bound may evolve as calculations are refined.
\par An additional indirect constraint on sterile-neutrino DM arises from comparing the observed number of Milky Way (MW) satellite galaxies to the results of $N$-body cosmological simulations. Compared to cold DM, warm DM particles are expected to suppress the matter power spectrum at small scales, reducing the number of low-mass DM subhaloes orbiting the Galaxy. In Fig.~\ref{fig:SNDM_NuSTAR_best}, we adopt the result of Ref.~\cite{Cherry:2017dwu} with $N_\text{subhalo}$\,=\,47, derived from SDSS data. Though a complete review of subhalo constraints on the properties of particle DM is beyond the scope of this paper, we note several important points. First, the Milky Way satellite population may not resemble that of a typical galaxy of its size and morphology, and surveys of dwarf galaxies targeting their stellar content must be corrected for completeness \cite{Kim:2018}. To address the former issue, surveys such as Satellites Around Galactic Analogues \cite{Geha:2017} aim to study the satellites of Milky Way analogues in the local Universe. Recent gravitational lensing surveys have also provided strong constraints on the properties of low-mass (down to ${\lesssim\,}10^8\,M_\odot$) subhaloes at cosmological redshifts unbiased by the haloes' stellar content \cite{Koopmans:2005a,Vegetti:2009a,Vegetti:2010a,Vegetti:2010b,Vegetti:2012,Vegetti:2017a,birrer:2017,Vegetti:2018a,Gilman:2019,Hsueh:2019,Gilman:2020}. In all of these cases, constraining $m_\chi$ using structure observables---both simulated and observed---also requires a model of the DM power spectrum, which is affected by its production mechanism, with all of the sources of uncertainty discussed in the previous paragraph \cite{Boyanovsky:2008nc,Boyanovsky:2010pw,Kuo:2018fgw,Menci:2018lis}.
\par Space-based x-ray observatories such as HEAO-1 \cite{10.1111/j.1365-2966.2006.10458.x}, \chandra{} \cite{RiemerSorensen:2009jp,Horiuchi:2013noa}, \xmm{} \cite{10.1111/j.1365-2966.2006.10458.x,Watson:2006qb,Malyshev:2014xqa,Iakubovskyi:2015dna}, \suzaku{} \cite{Loewenstein:2008yi,Tamura:2014mta}, \fermi{}-GBM \cite{Ng:2015gfa}, and \integral{} \cite{Yuksel:2007xh,Boyarsky:2007ge} have provided the most robust constraints on the $\chi \rightarrow \nu+\gamma$ decay rate for $m_\chi \simeq\text{1--100\,keV}$. The observation of an unknown x-ray line at $E \simeq 3.5\text{\,keV}$ (``the 3.5-keV line'') in several analyses \cite{Bulbul:2014sua,Boyarsky:2014jta,Boyarsky:2014ska} has led to much interest, as well as many follow-up analyses using different instruments and astrophysical targets \cite{Carlson_2015, Riemer-Sorensen:2014yda, Jeltema:2014qfa, Malyshev:2014xqa, Anderson:2014tza, Urban:2014yda, Tamura:2014mta, Sekiya:2015jsa, Figueroa-Feliciano:2015gwa, Riemer-Sorensen:2015kqa, Iakubovskyi:2015dna, Jeltema:2015mee, Ruchayskiy:2015onc, Franse:2016dln, Bulbul:2016yop, Hofmann:2016urz, Aharonian:2016gzq, Cappelluti:2017ywp,Boyarsky:2018ktr, Tamura:2018scp,Dessert:2018qih,Hofmann:2019ihc}. Some suggest that the 3.5-keV line may be a signature of sterile-neutrino DM~\cite{Abazajian:2014gza} or other DM candidates~\cite{Finkbeiner:2014sja, Higaki:2014zua, Brdar:2017wgy, Namjoo:2018oyn, Nakayama:2018yvj}; alternatively, modeling systematics \cite{Jeltema:2014qfa,Urban:2014yda} or novel astrophysical processes \cite{Gu:2015gqm,Gu:2017pjy} may play a role. Future high-spectral-resolution x-ray instruments may also be able to investigate the DM hypothesis for the origin of the 3.5-keV signal via velocity spectroscopy \cite{Speckhard:2015eva,Powell:2016zbo}. 
\par Since its launch in 2012, the \nustar{} observatory, due to its unique large-angle aperture for unfocused x-rays, has provided the leading constraints on sterile-neutrino DM across the mass range 10--50\,keV, leveraging observations of the Bullet Cluster \cite{Riemer-Sorensen:2015kqa}, blank-sky fields \cite{Neronov:2016wdd}, the Galactic center \cite{Perez:2016tcq}, and the M31 galaxy \cite{Ng:2019a}. In each of these cases, the \nustar{} observations were originally performed to study non-DM phenomena; therefore, DM searches using these data had to contend with large astrophysical backgrounds and/or reduced effective areas from masking bright point sources in the field of view (FOV). Improving upon these constraints, and extending them to the \nustar{} limit of $E = 3 \text{\,keV}$ (e.g., to test the tentative 3.5-keV signal), will therefore require observations with lower astrophysical backgrounds, as well as an improved model of the low-energy \nustar{} instrumental background.
\par In this paper, we present new constraints on the decay rate of sterile-neutrino DM particles using two \nustar{} observations, one ${\sim}10^\circ$ above and the other ${\sim}10^\circ$ below the Galactic plane, chosen to minimize astrophysical x-ray emission while still remaining near the center of the Galactic DM halo. These are the first \nustar{} observations dedicated to DM searches.
\newline 
\par In Sec.~\ref{sec:analysis}, we describe the data reduction and spectral modeling of the \nustar{} data, consistently incorporating the flux from the focused and unfocused FOVs. In Sec.~\ref{sec:dm_limits}, we combine the line flux limits from these new observations to constrain the $\chi \rightarrow \nu + \gamma$ decay rate for sterile neutrinos in the mass range 10--40\,keV, obtaining the strongest constraints to date in the 10--12\,keV mass range. We conclude in Sec.~\ref{sec:conclusions}.


\section{\label{sec:analysis} \nustar{} Data Analysis}
\par
In this section, we outline the aspects of the \nustar{} instrument that are relevant to our DM search, and describe \nustar{}'s unique wide-angle aperture for unfocused x-rays (Sec.~\ref{sec:instrument}). After describing the recent \nustar{} off-plane observations (Sec.~\ref{sec:newGC}) and our treatment of the \nustar{} instrument response (Sec.~\ref{sec:response}), we conclude with a discussion of the spectral model we use to analyze the data (Sec.~\ref{sec:spect_fit}).
\subsection{\label{sec:instrument} The \nustar{} Instrument}
\begin{table*}[t]
\centering
\caption{\nustar{} Galactic Bulge observations used in this analysis, with 0-bounce effective areas after data cleaning.}
\begin{tabularx}{\textwidth}{YYYYY}
\hline
\hline
NuSTAR obsID & Pointing (J2000) & Effective Exposure\footnote{After \texttt{OPTIMIZED} SAA filtering and manual data screening.}  & Detector Area $A_\text{0b}$\footnote{After bad pixel removal (both obsIDs) and point-source masking (40410001002 only).} & Solid Angle $\Delta \Omega_\text{0b}$\footnote{Average solid angle of sky for detecting 0-bounce photons, after correcting for bad pixel removal and vignetting efficiency.} \\
  & RA, Dec (deg) & FPMA / B (ks) & FPMA / B (cm$^2$) & FPMA / B (deg$^2$) \\
  \hline
  40410001002 & 253.2508,\: -26.6472  & 50.0 / 49.8 & {11.97 / 11.88} & {4.36 / 4.62} \\
 40410002002 &  280.3521,\: -27.6344  & 44.7 / 44.6 & {12.71 / 12.60}  & {4.53 / 4.56}  \\
 \hline
 \hline
\end{tabularx}
 \label{obs_table}
\end{table*}
\begin{figure*}[t!]
    \centering
    \vspace*{0.5cm}
    \hspace*{-1cm}
    \includegraphics[width=0.9\textwidth]{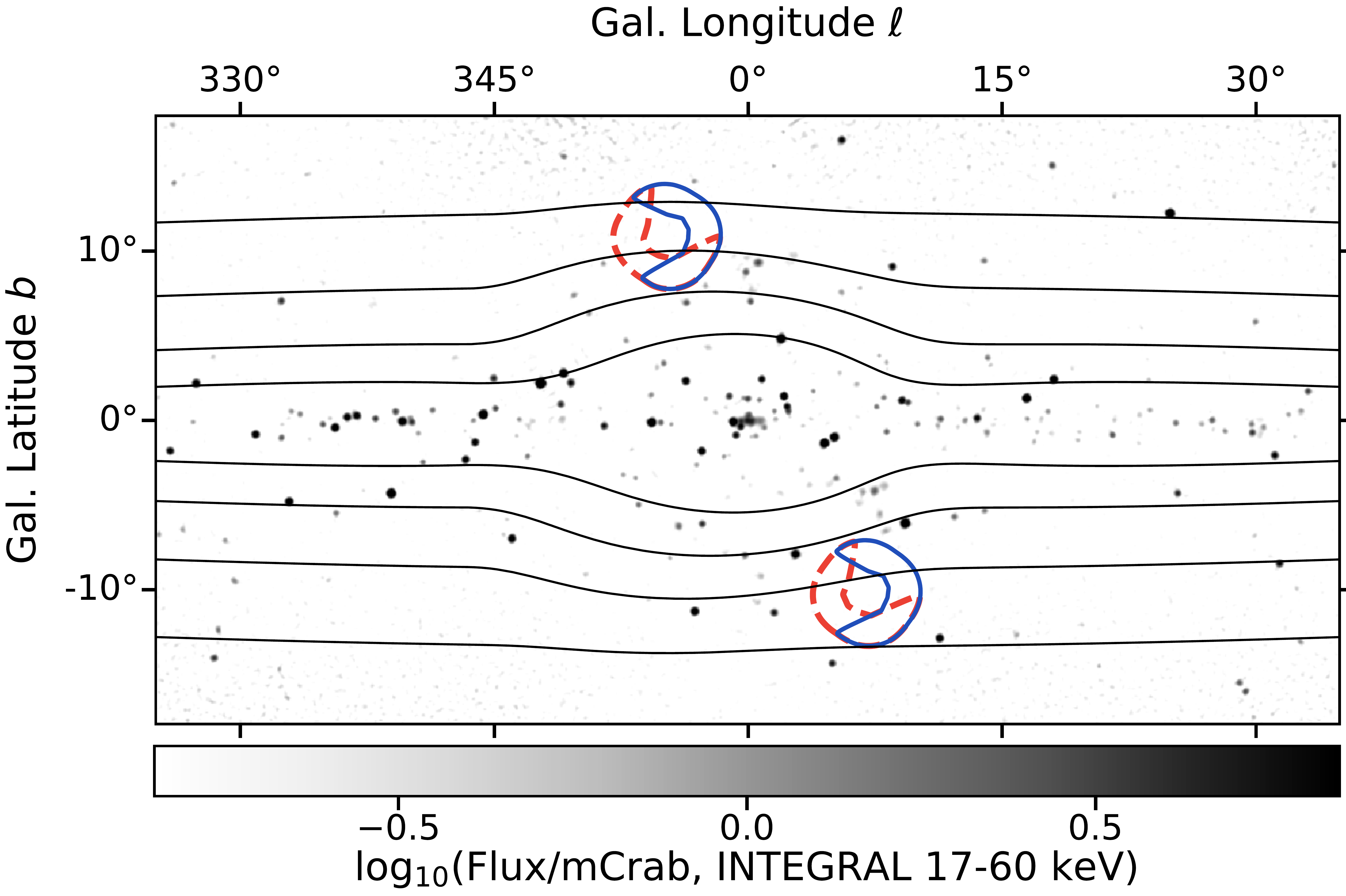}
    \caption{Sky map of the Galactic bulge region. The base color map shows the 17--60\,keV flux measured by INTEGRAL \cite{Krivonos:2017}, with many x-ray point sources clearly visible. The 0-bounce FOVs for the observations analyzed in this paper are indicated by the dashed red (FPMA) and solid blue (FPMB) ``Pac-Man''-shaped curves, and avoid known bright x-ray sources. The solid black contours indicate the predicted GRXE flux using the Galactic stellar mass model from Ref.~\cite{Launhardt:2002tx} and the GRXE emissivity model from Ref.~\cite{Revnivtsev:2006} (see Sec.~\ref{sec:spect_fit}). The contour values are symmetric about $b = 0^\circ$, decrease as $|b|$ increases, and are evenly spaced in $\log_{10} (\text{flux})$ between $10^{-12.5}$--$10^{-11}$ erg s$^{-1}$ cm$^{-2}$ deg$^{-2}$, inclusive.}
    \label{fig:GC_integral}
\end{figure*}
\par The \nustar{} instrument is more fully described in Refs.~\cite{Harrison:2013,Wik:2014,Madsen:2015}, with the aspects of the instrument relevant for our search technique described in our previous papers \cite{Perez:2016tcq,Ng:2019a}. Here, we summarize several key aspects. 
\par The \nustar{} instrument contains two identical, independent, and co-aligned telescopes, each consisting of a grazing-incidence Pt/C-coated x-ray optics module and a Focal Plane Module (FPM). The FPMs (labeled A and B) contain an aperture stop, a ${\sim}100\text{-}\mu\text{m}$ beryllium x-ray window with energy-dependent transmission efficiency $\mathcal{E}_\text{Be}(E)$, and a solid-state CdZnTe detector array with energy resolution ${\sim}0.4\text{\,keV}$ for x-rays with energies $E \lesssim 20\text{\,keV}$. Within the telescopes, properly-focused incoming x-rays reflect twice off the mirror segments, leading to their alternative name of 2-bounce (2b) photons. Both telescopes share essentially-overlapping $13^\prime\times 13^\prime$ FOVs for focused x-rays with energies between 3--79\,keV. The lower limit is primarily set by inactive material on the surface of the detector and $\mathcal{E}_\text{Be}(E)$ (see Secs.~\ref{sec:response} and~\ref{sec:spect_fit}), whereas the upper limit is set by the Pt K-edge of the mirror materials. The maximum x-ray energy recorded by the detectors is ${\sim}160\text{\,keV}$.
\par Unlike previous focusing x-ray telescopes such as \chandra{} or \xmm{}, the 10-m gap between the \nustar{} optics bench and the focal plane is open to the sky, allowing stray photons to strike the detector array without interacting with the mirror elements or being blocked by the aperture stops. For this reason, these unfocused x-rays are called 0-bounce (0b) photons. Although the 0-bounce effective area $A_\text{0b}$ is limited by the physical ${\sim}13\text{\,cm}^2$ area of each detector array, the effective 0-bounce FOV $\Delta \Omega_\text{0b}$ subtended by each array is ${\sim}4.5\text{\,deg}^2$, nearly two orders of magnitude larger than the 2-bounce FOV $\Delta \Omega_\text{2b}$, and more than counterbalancing the factor of ${\sim}20$ reduction in effective area between the 2-bounce and 0-bounce apertures. This approach provides a large increase in sensitivity to diffuse x-ray emission such as that expected from decaying DM in galactic halos, and thus the 0-bounce technique has been the dominant contribution to recent \nustar{} sterile-neutrino constraints \cite{Neronov:2016wdd,Perez:2016tcq,Ng:2019a}.

\subsection{\label{sec:newGC}\nustar{} Faint-Sky Off-Plane Observations}
The previous \nustar{} sterile-neutrino search in the Galactic center region \cite{Perez:2016tcq} was hampered by the presence of bright x-ray point sources in both the 0-bounce and 2-bounce FOVs, whose removal from the data greatly reduced the effective area, as well as a large continuum background from the Galactic ridge x-ray emission (GRXE, see Sec.~\ref{sec:spect_fit}) which was the dominant background component for $E \lesssim 20\text{\,keV}$. To combat both of these issues, we designed two dedicated \nustar{} observations (see Table~\ref{obs_table}), one ${\sim}10^\circ$ above the Galactic plane (obsID 40410001002), and the other ${\sim}10^\circ$ below (40410002002). The high Galactic latitude of these fields was chosen to minimize the GRXE continuum background while still remaining near the center of the Galactic DM halo, as well as avoiding known bright x-ray sources near the Galactic plane (see Fig.~\ref{fig:GC_integral}).
\par The \nustar{} observations described above were carried out in August and October 2018, with an initial unfiltered exposure time of ${\sim}200\text{\,ks}$ (summed over both obsIDs and FPMs). Data reduction and analysis are performed using the \nustar{} Data Analysis Software pipeline, \textsc{nustardas v1.5.1}. The flags \texttt{SAAMODE=OPTIMIZED} and \texttt{TENTACLE=YES} are used to remove events coincident with \nustar{} passages through the South Atlantic Anomaly (SAA), and ``bad pixels'' (defined in the \nustar{} calibration database) are removed. We observe a faint x-ray point source near the edge of the 2-bounce FOV in obsID 40410001002, whose position is consistent with the chromospherically-active stellar binary HD 152178 \cite{Gaia:2016,Gaia:2018b}. This system has also been detected in x-rays by RXTE \cite{Voges:1999ju} and \suzaku{} \cite{Mori:2012}. To eliminate systematic uncertainties associated with modeling this source's spectrum, we remove from our analysis all x-ray events in a circular region of radius $75^{\prime\prime}$ around the nominal position of the source in both FPMs, excluding ${\gtrsim}80\%$ of the source photons \cite{An:2014hua}. (The position of the x-ray source 1RXS J165306.1-263434 also lies within the 2-bounce FOV of this obsID \cite{Voges:1999ju}; however, it is sufficiently faint that its \nustar{} spectrum is consistent with background, so we do not exclude it from the analysis. There are no x-ray point sources visible in obsID 40410002002.) Finally, we inspect the $\text{3--10\,keV}$ light-curves of each observation to check for transient fluctuations due to solar activity or unfiltered SAA events, and remove any time intervals with a count rate ${>}2.5\sigma$ from the quiescent average. After all cuts, the total cleaned exposure time used in this analysis, summed over both obsIDs and telescopes, is ${\sim}190\text{\,ks}$.
\par We extract spectra from the full detector planes as extended sources using the \textsc{nuproducts} routine in \textsc{nustardas}, and bin each spectrum with equal logarithmic separations $\Delta \log_{10}E = 0.01$ (i.e., 100 bins per decade) in the energy ranges $\text{5--20\,keV}$ and $\text{95--110\,keV}$. This provides a statistical uncertainty that is everywhere ${\sim}10\%$ per bin while also being narrower than the ${\sim}0.4\text{-keV}$ \nustar{} energy resolution across the energy range $\text{5--20\,keV}$. As described in Ref.~\cite{Ng:2019a}, we exclude the energy range $\text{3--5\,keV}$, as the behavior of the low-energy \nustar{} background---particularly the origin of the 3.5- and 4.5-keV lines in the default background model---is the subject of active investigation. (Additionally, including the $\text{3--5\,keV}$ region can bias the determination of the internal power-law parameters discussed in Sec.~\ref{sec:spect_fit}; see Ref.~\cite{Ng:2019a} for details.) We also exclude the energy range $\text{20--95\,keV}$, as this region is dominated by a forest of instrumental lines. DM constraints in this energy range are therefore weakened and prone to systematic effects, as discussed in Refs.~\cite{Neronov:2016wdd,Perez:2016tcq,Ng:2019a}. Excluding this energy range also speeds up our analysis, and we verify that it does not affect our results in the $\text{5--20\,keV}$ energy range. Finally, we note that the $\text{20--95\,keV}$ energy range has already been largely excluded by previous sterile-neutrino searches using data from Fermi-GBM \cite{Ng:2015gfa}, INTEGRAL \cite{Boyarsky:2007ge}, and \nustar{} \cite{Neronov:2016wdd,Perez:2016tcq,Ng:2019a}.

\subsection{\label{sec:response}\nustar{} Response Files}
\par To describe the effects of the detector effective area and solid angle for the CXB, GRXE, and DM line components described in Sec.~\ref{sec:spect_fit}, we define custom response files that relate the measured event rate $d^2N/dEdt$ to the astrophysical flux. For 0-bounce components, the response is $\mathcal{E}_\text{Be}(E) A_\text{0b}\Delta \Omega_\text{0b}$, where the grasp $A_\text{0b}\Delta\Omega_\text{0b}$ is calculated using the \textsc{nuskybgd} code \cite{Wik:2014} and $\mathcal{E}_\text{Be}(E)$ is the Be window transmission efficiency. For 2-bounce components, the response is $\mathcal{E}_\text{Be}(E)A_\text{2b}(E)\Delta\Omega_\text{2b}$, where $\mathcal{E}_\text{Be}(E)$ and $A_\text{2b}(E)$ are calculated by \textsc{nustardas}, extracting the entire FOV as an extended source using $\text{\textsc{nuproducts}}$. Here, $\Delta \Omega_\text{2b}$ is simply the geometric area of the 2-bounce FOV, and is ${\sim}0.046\text{\,deg}^2$ for obsID 40410001002 and ${\sim}0.047\text{\,deg}^2$ for obsID 40410002002, the former being slightly less than the latter due to the exclusion of the $75^{\prime\prime}$-radius circle around the point source. The responses for internal detector components---the internal continuum, power-law, and lines---are calculated by $\text{\textsc{nuproducts}}$, and do not depend on area or solid angle.

\subsection{\label{sec:spect_fit} \nustar{} Spectral Modeling}
\begin{table*}[t]
\centering
\caption{The \nustar{} spectral model used in this paper. Parameters with numerical values are frozen to those values, and all free parameters are allowed to vary independently between FPMA/B and between the two obsIDs.}
\begin{tabularx}{\textwidth}{CCCD}
\hline
\hline
& & & \\
 Model component & \textsc{xspec} model\footnote{The CXB, GRXE, and DM line models also include absorption from the interstellar medium through the \texttt{tbabs} model with fixed column density $N_\text{H}$, as well as absorption from the beryllium x-ray shield. All model components except the internal continuum include the absorption effects of detector surface material. See Sec.~\ref{sec:spect_fit} for details.} &  Parameter  & Value \\
& & & \\
\hline \\
 CXB & \texttt{powerlaw}*\texttt{highecut} & 3--20 keV flux & $2.6\times 10^{-11}\text{\,erg\,s}^{-1}\text{\,cm}^{-2}\text{\,deg}^{-2}$ \cite{Gruber:1999,Churazov:2006bk}
 \\
  & & Spectral index $\Gamma$ & $1.29$ \cite{Gruber:1999,Churazov:2006bk} \\
  & & $E_\text{cut}$ & $10^{-4}\text{\,keV}$ \\
  & & $E_\text{fold}$ & $40$ keV \cite{Gruber:1999,Churazov:2006bk} \\
  & & &  \\
 GRXE & \texttt{apec} & 3--20 keV flux & Free \\
 & & Plasma $kT$ & 8 keV \cite{Kaneda:1997,Yuasa:2012,Perez:2019a} \\
 & & Abundance ratio & Free within 0--1.2 \\
 & & & \\
 Internal continuum & \texttt{bknpower} & $E_\text{break}$ & 124 keV \cite{Wik:2014} \\
 & & $\Gamma(E < E_\text{break})$ & $-0.05$ \cite{Wik:2014}\\
 & & $\Gamma(E > E_\text{break})$ & $-0.85$ \cite{Wik:2014} \\
 & & Normalization & Free \\
 & & & \\
 Internal power-law & \texttt{powerlaw} & Spectral index $\Gamma$ & Frozen for each FPM/obsID (Sec.~\ref{sec:spect_fit}) \\
 & & Relative norm. & Frozen for each FPM/obsID (Sec.~\ref{sec:spect_fit})\\
 & & & \\
 Internal lines & \texttt{lorentz} & Line energies & 10.2, 19.7, 104.5 keV \cite{Wik:2014}\\
 & & Line widths & 0.6, 0.2, 0.5 keV \cite{Wik:2014}\\
 & & Line norms. & Free \\
 & & & \\
 DM line & \texttt{gaussian} & Line energy & See Sec.~\ref{sec:spect_fit} \\
  & & Line width & 0 keV \\
 & & Line flux & See Sec.~\ref{sec:spect_fit} \\
 \hline
 \hline
\end{tabularx}
 \label{model_table}
\end{table*}

\begin{figure*}
    \centering
    \includegraphics[width=\textwidth]{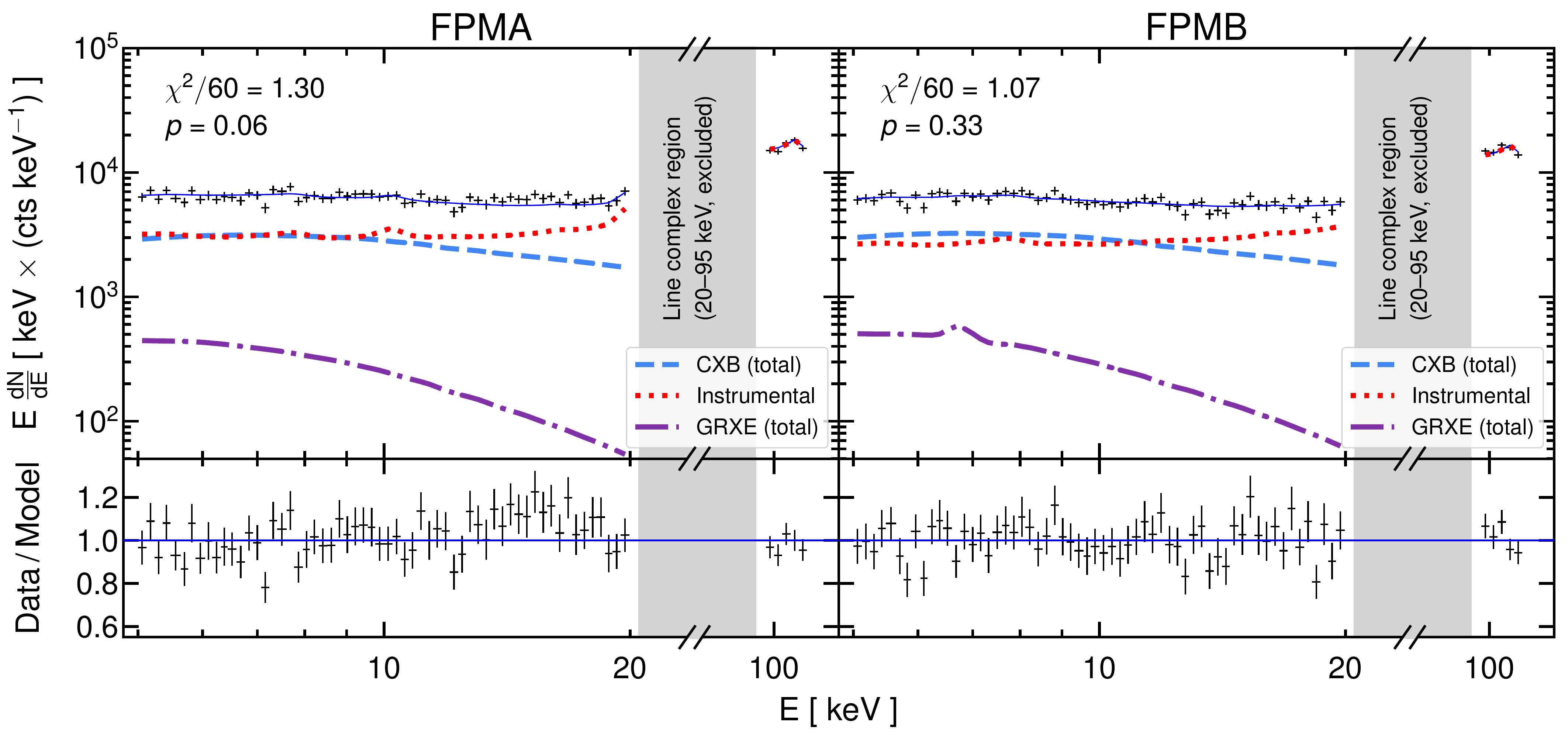}
    \caption{Data and model spectra for obsID 40410001002, with FPMA (left) and FPMB (right), including contributions from the CXB, instrumental background, and the GRXE. The error bars correspond to $\pm 1\sigma$ statistical uncertainties, and the CXB and GRXE curves incorporate both 0-bounce and 2-bounce emission. We exclude the energy range 20--95\,keV as it is dominated by internal detector lines (in previous analyses \cite{Perez:2016tcq,Ng:2019a}, we have already probed this range well), though we include the energy range 95--110\,keV to constrain the internal detector continuum. See Sec.~\ref{sec:spect_fit} for details.}
    \label{fig:1002}
\end{figure*}

\begin{figure*}
    \centering
    \vspace*{2cm}
    \includegraphics[width=\textwidth]{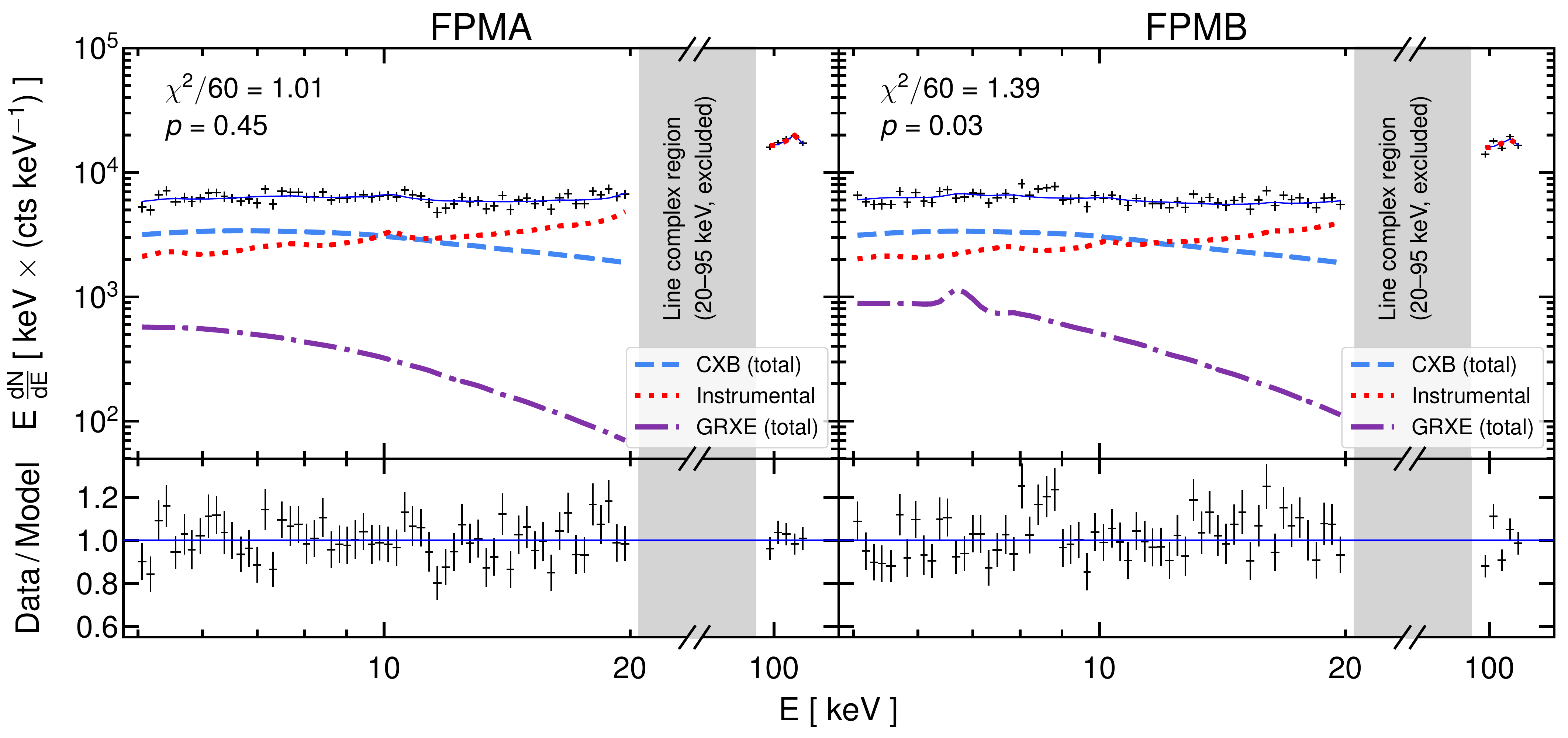}
    \caption{Same as Fig.~\ref{fig:1002}, but for obsID 40410002002.}
    \label{fig:2002}
\end{figure*}

\par Our spectral model contains six components, which may be broadly classified as having instrumental or astrophysical origins (see Table~\ref{model_table}). The instrumental background consists of a low-energy internal power-law dominant at energies $E \lesssim 10 \text{\,keV}$, the internal detector continuum, and a series of phenomenologically-motivated lines. The astrophysical components include the cosmic x-ray background (CXB), with an event rate similar to the instrumental components' over the energy range of this analysis; and the GRXE, whose flux is a factor ${\sim}10$ lower than the CXB. The treatment of each of these model components is described in this section.
\par To describe the internal continuum and line backgrounds, we adopt the default \nustar{} spectral model of Ref.~\cite{Wik:2014}. The internal continuum is parameterized by a broken power-law with $E_\text{break} = 124\text{\,keV}$, and the line energies and widths are frozen to the values in the default model, with only the line normalizations free to fit. (The 124-keV break is outside the energy range of our analysis, and thus does not affect the fit; we include it merely for continuity with the default \nustar{} model.) The line normalizations are also allowed to vary between each of the spectra, accounting for differences in the instrumental background conditions between the FPMs. We retain the 95--110\,keV data as the event rate in this range is dominated by the internal continuum, and is necessary to constrain the overall continuum normalization. We explore alternative high-energy intervals with endpoints around 95\,keV and 120\,keV, and find that the fit quality is not sensitive to the precise values of the endpoints, provided the interval is sufficiently wide to constrain the internal continuum.
\par The default \nustar{} instrumental background model \cite{Wik:2014} includes a ${\sim}1$-keV collisionally-ionized plasma component (the \texttt{apec} model in \textsc{xspec} \cite{Smith_2001}) which is strongest for energies ${E < 5\text{\,keV}}$ and is believed to result from reflected solar x-rays. Unfortunately, this model provides a poor fit ($\chi^2/\text{d.o.f.}\gtrsim 1.7$) to the observed spectrum, with the residuals indicating a clear excess in the energy range 5--10\,keV. As we exclude the ${E < 5\text{\,keV}}$ data, we adopt the procedure described in Refs.~\cite{Ng:2019a,Perez:2019a} and replace the \texttt{apec} model with a power-law. For each FPM and each obsID, we use the data collected when the telescope aperture is occulted by the Earth to constrain the power-law spectral index and normalization with respect to the internal continuum. As the Earth completely fills the 0-bounce and 2-bounce apertures during occultation mode, we assume that the astrophysical components contribute negligible flux, and include only the internal detector components when modeling the occulted data. In particular, the Earth albedo flux is suppressed by at least one order of magnitude compared to the CXB and GRXE over the energy range of our analysis \cite{Sazonov:2007,Churazov:2008,Fioretti:2012}. The spectral index and relative normalization of the internal power-law are frozen to their best-fit occultation-mode values during fits to the science data, shown in Figs.~\ref{fig:1002} and~\ref{fig:2002}. This procedure provides a much better fit (${\chi^2/\text{d.o.f.}\lesssim 1.4}$) to the observed science-mode spectra over the energy range of our analysis; however, there are still noticeable deviations, which will be discussed later in this section, and in Sec.~\ref{sec:monte_carlo}.
\par The cosmic x-ray background (CXB) arises from unresolved extragalactic sources, and constitutes one of the dominant irreducible \nustar{} backgrounds in both the 0-bounce and 2-bounce FOVs. As specified in the default \nustar{} background model, we parameterize the CXB spectrum with a cut-off power-law whose flux, spectral index, and $e$-folding energy are fixed to the values measured in similar energy ranges by HEAO-1 and INTEGRAL \cite{Gruber:1999,Churazov:2006bk}; i.e., there are no free parameters in the CXB model. This choice is supported by a previous \nustar{} analysis using the 0-bounce technique, which obtained a CXB flux consistent with our adopted value \cite{Perez:2019a}. We test the effect of allowing the CXB flux to vary by $\pm10\%$ to account for cross-calibration uncertainty or the effects of cosmic variance in the ${\sim}4.5\text{\,deg}^2$ FOV, as the number density of CXB sources was previously measured by \nustar{} to be ${\gtrsim}100\text{\,deg}^{-2}$ \cite{Harrison:2015yuk}. We find no significant change in the fit quality. Similarly, we examine the effects of allowing the CXB spectral index to be unconstrained. In three of the spectra the best-fit CXB spectral index is consistent with our adopted value at ${{>}90\%}$ confidence, whereas in spectrum 40410001002A the best-fit value is ${{<}\,1}$ (inconsistent with previous measurements by HEAO-1 and INTEGRAL). The fit quality is not significantly improved by allowing the CXB spectral index to vary in any of our spectra, so we fix it to the value in Table~\ref{model_table}. Finally, the \texttt{highecut} term brings a factor $\exp[(E_\text{cut}-E)/E_\text{fold}]$ for $E \ge E_\text{cut}$ and is constant for $E\le E_\text{cut}$, so we choose $E_\text{cut} = 10^{-4}\text{\,keV}$ to ensure that the exponential folding is applied over the full energy range of our analysis. As shown in Figs.~\ref{fig:1002}  and~\ref{fig:2002}, the CXB is the dominant astrophysical background in these off-plane observations.

\par The GRXE is believed to result from unresolved point sources in the Galactic ridge \cite{Revnivtsev:2009mf}, and its emissivity is observed to trace the near-infrared surface brightness (and hence stellar density) of the Galaxy \cite{Revnivtsev:2006,Krivonos:2012,Revnivtsev:2012,Yuasa:2012}. Broadband studies of the GRXE indicate that it is likely a multi-temperature plasma, with $kT_1\lesssim 1\text{\,keV}$ and $kT_2 \sim 8\text{\,keV}$ \cite{Kaneda:1997,Yuasa:2012}. We model the GRXE, which appears in both the 0-bounce and 2-bounce FOVs, as a single-temperature collisionally-ionized plasma (the \texttt{apec} model described previously) with a fixed temperature of 8\,keV previously measured by \nustar{}; however, this analysis was not sensitive to the elemental abundances \cite{Perez:2019a}. (We are unable to leave the GRXE temperature free to fit, as we find that doing so leaves the temperature almost completely unconstrained.) Particularly strong emission lines between 6--7\,keV arise from K$\alpha$ transitions in neutral and highly-ionized Fe, and it was these lines which limited the sensitivity of the previous \nustar{} sterile-neutrino search near the Galactic center (see Ref.~\cite{Perez:2016tcq} and Fig.~\ref{fig:dmgamma_both_figs} of this paper).
\par It is important to note that the ``GRXE'' component in our spectral model includes flux from the GRXE, un-modeled point sources, reflected x-rays from the Earth's atmosphere, and any low-energy instrumental backgrounds not described by our default spectral model, as the GRXE component includes the only free normalization parameter in the low-energy part of our spectral model. Therefore, we leave both the GRXE elemental abundance (as a ratio to solar) and flux as free parameters, where the flux is unconstrained and the abundance ratio is constrained to the range 0--1.2. The 0-bounce and 2-bounce GRXE components are constrained to have the same flux and abundance ratio.
\par The lower bound on the GRXE abundance ratio arises from the requirement that elemental abundances be strictly positive, and the upper bound is motivated by previous measurements of the GRXE \cite{Yuasa:2012}. Additionally, freezing the abundance ratio to a nonzero value can force the GRXE flux to unreasonable extremes as the model attempts to fit the GRXE by way of its emission lines, thereby biasing the rest of the 5--20\,keV fit. The flux of the GRXE emission lines is directly related to the number of atoms in the FOV undergoing electronic de-excitation, and hence to the elemental abundances of the plasma; as shown by the slight bump in Figs.~\ref{fig:1002} and~\ref{fig:2002}, the fits to the FPMB spectra of both obsIDs prefer a slightly higher GRXE abundance ratio than the FPMA spectra, though this difference is within the uncertainty on the value of the abundance parameter. 
\par Finally, the freedom in the GRXE flux acts to account for any un-modeled CXB flux, as the two components have similar continuum shapes in the ${E < 10\text{\,keV}}$ range, where their flux is highest. By fixing the CXB and allowing the GRXE flux to float, we consistently account for any variance in the flux of both components, and we find that the best-fit GRXE flux is consistent with Galactic stellar mass and emissivity models \cite{Launhardt:2002tx,Revnivtsev:2009mf}. Additionally, we find that allowing both the CXB and GRXE fluxes to vary leads to best-fit values which are inconsistent with the previously-described measurements of these components' flux levels.
\par We parameterize our DM line signal in \textsc{xspec} with a vanishingly-narrow Gaussian---i.e., a $\delta$-function in $E$---as the intrinsic width of any DM line is expected to be much less than the ${\sim}0.4$\,keV detector energy resolution with which it is convolved. Our treatment of the DM line during the line-search procedure is described further in Sec.~\ref{sec:dm_search}.
\par The fluxes of the astrophysical components in our spectral model---CXB, GRXE, and DM line---are attenuated by absorption and scattering in the interstellar medium (ISM). This attenuation is parameterized in terms of the equivalent column density of neutral hydrogen, $N_\text{H}$, via the \texttt{tbabs} model in \textsc{xspec} \cite{Wilms:2000ez}. We adopt fixed values of $7.0\times 10^{20}\text{\,cm}^{-2}$ for obsID 40410001002 and $1.1\times 10^{21}\text{\,cm}^{-2}$ for obsID 40410002002 \cite{Dickey:1990mf,Kalberla:2005ts}. (Both FPMs share the same $N_\text{H}$ value, which is assumed to be constant across the 0-bounce and 2-bounce FOVs despite the somewhat different sky coverage and values of $\Delta\Omega_\text{0b}$ from A/B.) This corresponds to
an optical depth $\tau \lesssim 10^{-2}$ at $E = 5\text{\,keV}$, falling steeply with increasing energy. Although the flux attenuation from the ISM is a $\lesssim 1\%$ effect across the energy range of this analysis, we include it for consistency.
\par Finally, we consider the absorption of x-rays within the \nustar{} instrument itself. Before incoming astrophysical x-rays (from the CXB, GRXE, or DM) strike the detectors, they must pass through a ${\sim}100\text{-}\mu\text{m}$ beryllium shield with transmission efficiency $\mathcal{E}_\text{Be}(E)$, rising from ${\sim}0.67$ at $E=3$\,keV to ${\sim}0.92$ at $E=5$\,keV. (The treatment of $\mathcal{E}_\text{Be}$ is discussed further in Sec.~\ref{sec:response}.) An additional absorption effect arises in the detectors themselves. The CdZnTe detectors have a ${\sim}0.11\text{-}\mu\text{m}$ Pt contact coating, as well as a ${\sim}0.27\text{-}\mu\text{m}$ layer of inactive CdZnTe (both varying somewhat between individual detector crystals), through which incoming x-rays must pass \cite{Madsen:2015}. At $E=5$\,keV, these detector components result in a flux attenuation of ${\sim}25\%$, though this decreases quickly with increasing energy \cite{Ng:2019a}. These detector absorption effects (often called \texttt{nuabs} or \texttt{detabs}) are included in every spectral component except the internal continuum.
\par As shown in Figs.~\ref{fig:1002} and~\ref{fig:2002}, the model described in Sec.~\ref{sec:spect_fit} provides an acceptable fit to the \nustar{} spectra across most of the 5--20\,keV energy range (see Figs.~\ref{fig:1002} and \ref{fig:2002} for the reduced-$\chi^2$ and corresponding $p$-values for each spectrum), but there are several deviations from the model that may affect our derived line flux limits, and thus require further consideration.
The higher $\chi^2$ in FPMA of obsID 40410001002A is due to the energy range 15--20\,keV (excluding this energy range yields ${\chi^2 / 47 = 0.94}$ with $p = 0.59$), and similarly for FPMB of obsID 40410002002B in the energy range 8--9\,keV (yielding $\chi^2/54 = 1.15$ with $p = 0.21$). As both of these regions are excesses with respect to the default background model, the DM line flux limits in the mass ranges ${m_\chi \simeq 16\text{\,keV}}$ and 30--40\,keV are correspondingly weakened (see Sec.~\ref{sec:monte_carlo}), as we use a conservative line-search procedure in which the DM line flux is allowed to fill the excess (see Sec.~\ref{sec:dm_search}). 
In Sec.~\ref{sec:monte_carlo}, we perform Monte Carlo simulations to verify that our constraint is consistent with one limited by statistical variations in our measurement, not systematic variations due to incomplete modeling.


\section{\label{sec:dm_limits} \nustar{} DM Analysis}
\par In this section, we describe the procedure used to search for DM line signals and set upper limits on the decay rate of DM to final states including a single mono-energetic photon (Sec.~\ref{sec:dm_search}), and compare to sensitivity estimates from simulations (Sec.~\ref{sec:monte_carlo}). Finally, we discuss the implications for sterile-neutrino dark matter (Sec.~\ref{sec:constraints}).

\subsection{DM Line Search \label{sec:dm_search}  }
\begin{figure*}
\centering
\begin{minipage}[b]{.48\textwidth}  
    \includegraphics[scale=0.395,right]{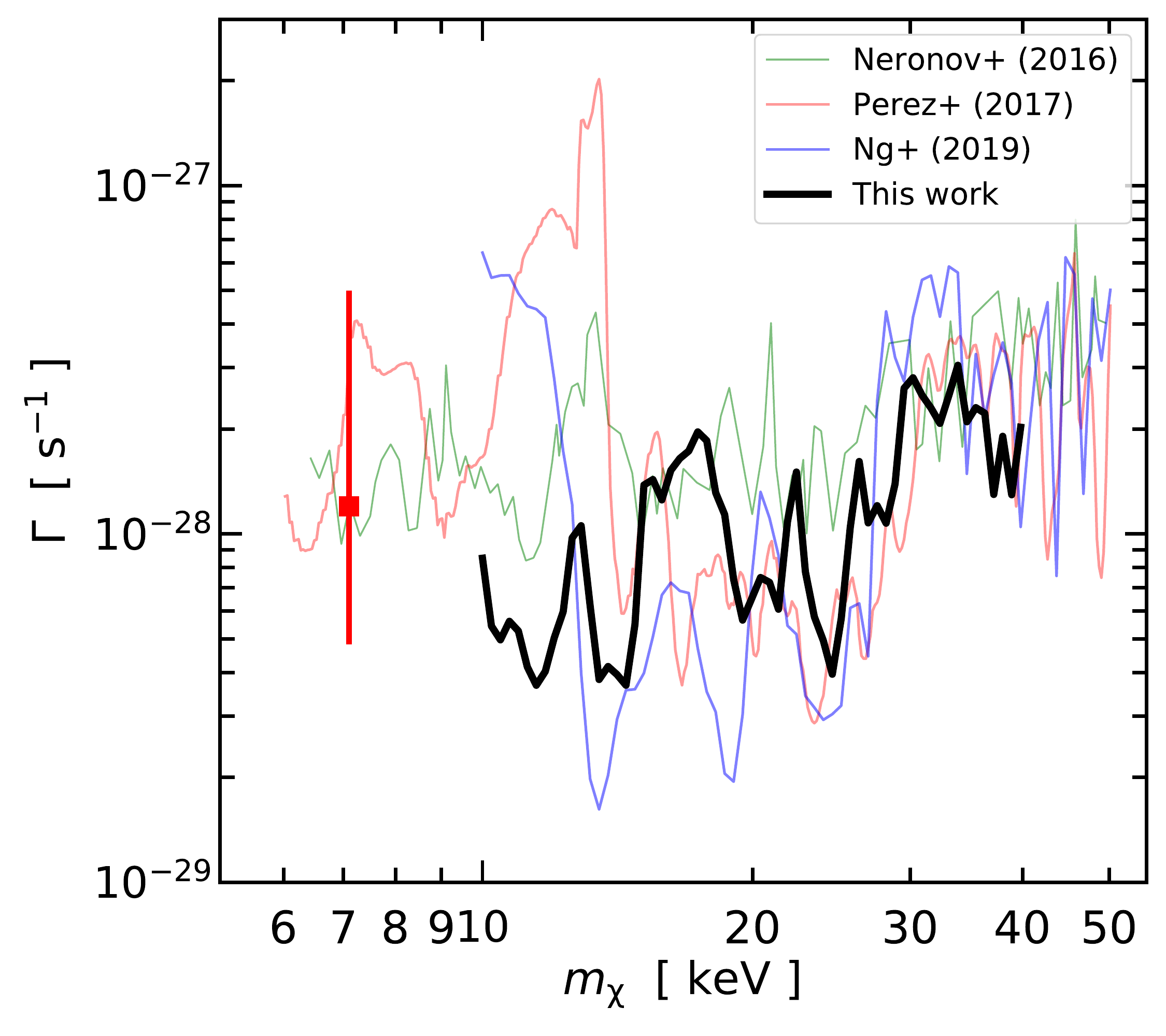}
\end{minipage}\qquad
\begin{minipage}[b]{.48\textwidth}
    \includegraphics[scale=0.395,right]{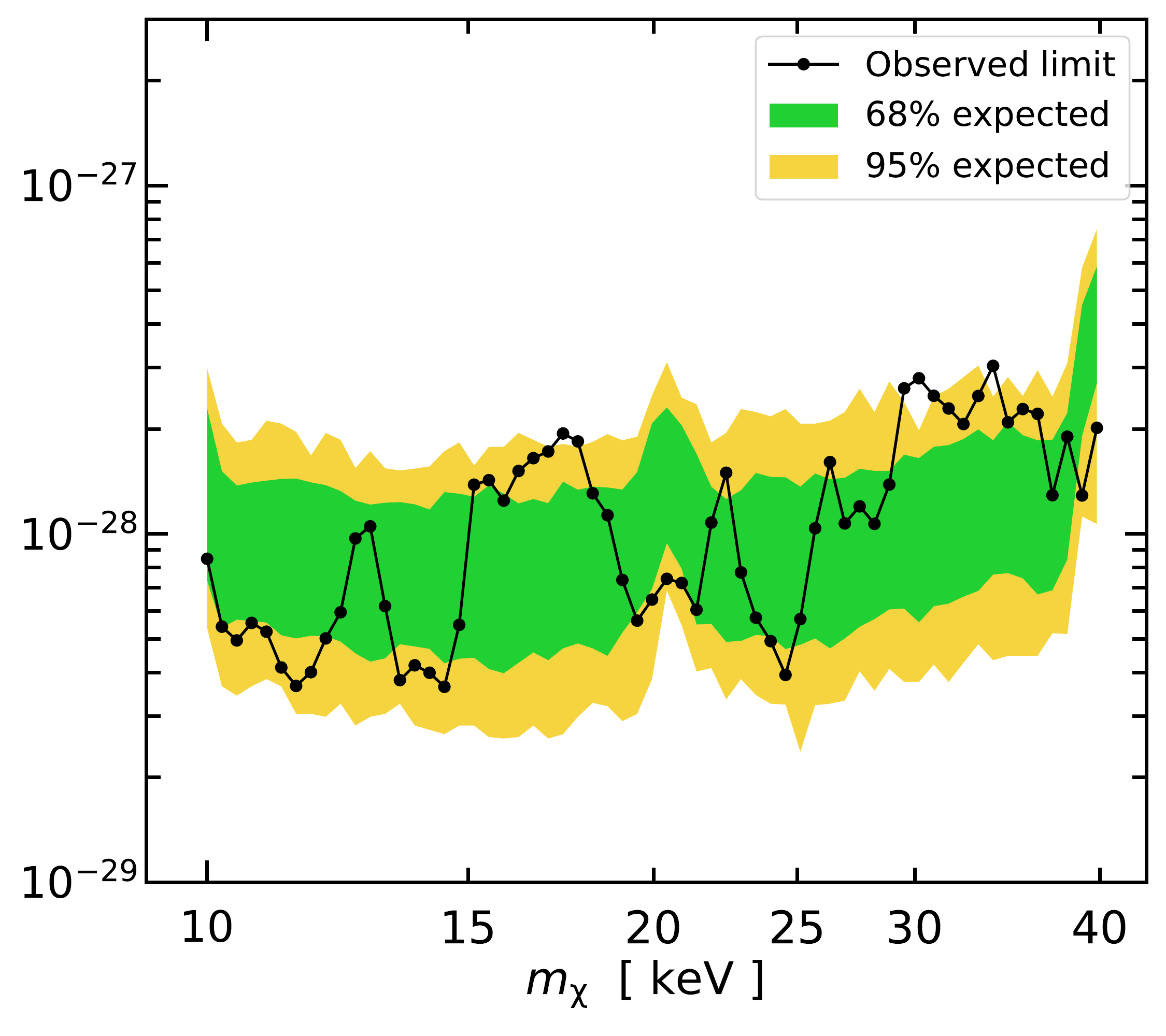}
\end{minipage}
\caption{ \textbf{Left:} Comparison of the limit obtained in this paper to that from several surveys using the 0-bounce technique, including blank sky (green, Ref.~\cite{Neronov:2016wdd}), Galactic center (red, Ref.~\cite{Perez:2016tcq}), and M31 fields (blue, Ref.~\cite{Ng:2019a}), as well as the tentative signal at $E\simeq 3.5\text{\,keV}$ (red point, Refs.~\cite{Bulbul:2014sua,Boyarsky:2014jta,Boyarsky:2014ska}).  With only ${\sim}$190\,ks, we have achieved comparable constraints to analyses with much deeper exposures \cite{Neronov:2016wdd,Perez:2016tcq,Ng:2019a}. We have achieved the best constraint in 10--12\,keV mass range, essential for investigating the remaining $\nu\text{MSM}$ parameter space shown in Fig.~\ref{fig:SNDM_NuSTAR_best}.
\textbf{Right:} The observed 95\% upper limit on the DM decay rate $\Gamma$ obtained in this paper, compared to the expected 68\% (green) and 95\% (yellow) sensitivity bands from simulations (see Sec.~\ref{sec:monte_carlo}).}
\label{fig:dmgamma_both_figs}
\end{figure*}

Equipped with the spectral model described in Sec.~\ref{sec:spect_fit}, we search for DM line signals in the two observations.  Our search procedure follows closely that from Refs.~\cite{Perez:2016tcq, Ng:2019a}, and is briefly described here. 

We divide the 10--40\,keV mass band into bins with equal logarithmic separations $\Delta\log_{10}m_{\chi} = 0.01$ (i.e., 100 bins per decade in $m_\chi$). At each mass bin, we add a DM line with photon energy $E = m_\chi/2$ to the model. The number of DM photons in the line for each module and observation is 
\begin{equation}
	N_{\rm DM} = \frac{\Gamma}{4 \pi m_{\chi}} T\, A_{\rm 0b}\, \Delta\Omega_{\rm 0b}\, {\cal J}(1+ f_{\rm 2b}), \,
\end{equation}
where $\Gamma$ is the decay rate, $m_{\chi}$ is the DM mass, $T$ is the observation time, $A_{\rm 0b}$ and $\Delta\Omega_{\rm 0b}$ are the 0-bounce effective area and effective FOV defined in Sec.~\ref{sec:response}, ${\cal J}$ is the FOV-averaged line-of-sight integral of the DM density~(J-factor), and $f_{\rm 2b}$ is the energy-dependent contribution from the 2-bounce component~(see Fig. 3 in Ref.~\cite{Ng:2019a} for the energy dependence of the 2-bounce contribution; in this work, we find a modest ${\sim}20\%$ enhancement at $E = 10\text{\,keV}$).

 To obtain the J-factors, we consider several DM density profiles. One popular choice is the generalized Navarro-Frenk-White (NFW) profile, ${\rho \propto (r/r_{s})^{-\gamma} (1+ r/r_{s})^{\gamma -3}}$. For the standard (DM-only) NFW profile, we adopt an inner slope ${\gamma = 1}$ and scale radius ${r_{s} = 20\,{\rm kpc}}$ \cite{Navarro:1996gj,Hooper:2016ggc}.  We fix the galactocentric solar radius and the local DM density to be $8\,{\rm kpc}$ and $0.4\,{\rm GeV\,cm^{-3}}$, respectively ~\cite{Pato:2015dua,Boehle:2016a,Gravity:2019a}. The standard NFW profile was found to be a good fit to the Milky Way kinematic data~\cite{Nesti:2013uwa}, but it has been suggested that the density profiles could be flattened below 1.5\,kpc~\cite{Calore:2015oya, Schaller:2015mua}. Therefore, we also consider the more conservative choice, coreNFW, where we set a density core below 1.5\,kpc---i.e., $\rho(r\,{<}\,1.5\,{\rm kpc})$\,=\,$\rho(1.5\,{\rm kpc})$.  Another conservative NFW variant we consider is the sNFW, where we use a shallower index ${\gamma=0.7}$~\cite{Pato:2015dua}. 
We use sNFW as our default result, obtaining ${{\cal J} \simeq 20\,{\rm GeV\,cm^{-3}\,kpc\,sr^{-1}}}$ for the observation regions in this analysis; for NFW and coreNFW, the J-factors are larger by ${\sim}20\%$.

Another shallow density profile often considered in the literature is the Burkert profile ${\rho \propto  (1+r/r_{s})^{-1}(1+ r^{2}/r_{s}^{2})^{-1}}$ ~\cite{Burkert:1995yz}, with best-fit local DM density ${\sim}0.5\,{\rm GeV\,cm^{-3}}$ and scale radius ${{\sim}8\text{\,kpc}}$ \cite{Nesti:2013uwa,Lin:2019a}. This profile effectively has a density core within $r_{s}$, which we note is much larger than what was found in Refs.~\cite{Calore:2015oya,Schaller:2015mua}. Even in this case, the J-factor is only ${\sim}10\%$ smaller than our default sNFW choice. This small deviation shows the robustness of our results, and reflects an additional advantage of using observations slightly offset from the Galactic center.

At each DM mass, the only free parameter for the DM line is the decay rate. We find the best-fit $\chi^{2}(\Gamma)$ distribution for each module and observation by scanning through a range of $\Gamma$, refitting the entire spectral model to find the minimum $\chi^{2}$ value for each $\Gamma$. This line-search procedure is conservative, as it allows the DM line to attain the full strength of any background lines.
\par The sensitivity of the two observations~(four separate fits including both modules) at each $m_\chi$ are combined by adding the respective $\chi^{2}$ distributions:
\begin{equation}
	X^{2}(\Gamma) = \sum_\text{obs} \chi^{2}(\Gamma) \, . 
\end{equation}
We note that for each module, the background parameters are allowed to be independent~(see Sec.~\ref{sec:spect_fit} for exceptions). Compared with simply stacking the spectra, this combining procedure is used to avoid potential systematic errors due to combining observations with different instrumental and/or astrophysical backgrounds.  

\par The minimum in $X^2(\Gamma)$ for each mass bin corresponds to the best-fit decay rate $\Gamma_{\rm min}$, with a $5\sigma$ line detection requiring $X^2(\Gamma_{\rm min}) - X^2(\Gamma = 0) < -25$. We find no signals consistent with decaying DM in the mass range 10--40\,keV, and instead set upper limits on the DM decay rate. The 95\% one-sided upper limit, $\Gamma_{95}$, occurs at $X^2(\Gamma_{95}) = X^2(\Gamma_{\rm min}) + 2.71$, and is shown in both frames of Fig.~\ref{fig:dmgamma_both_figs}. In the 10--40\,keV mass range, our results are comparable to previous \nustar{} limits from blank-sky~\cite{Neronov:2016wdd}, Galactic center~\cite{Perez:2016tcq}, and M31 observations~\cite{Ng:2019a}. In particular, we are able to improve upon previous constraints in the 10--12\,keV mass range by a factor of ${\sim}2$. Finally, we note that with only ${\sim}190\text{\,ks}$ exposure, our dedicated Galactic bulge observations are able to achieve sensitivity comparable with searches using several Ms combined exposure. This is due to the low astrophysical background, as well as the large J-factors in the chosen FOVs. 

\subsection{Sensitivity Estimation with Simulations \label{sec:monte_carlo}  }
\par To validate our results, we perform line searches in mock spectra to find the expected upper limits when the spectra are purely statistically limited. This exercise also allows us to further study the deviations discussed in Sec.~\ref{sec:spect_fit}.
\par Instead of fully mimicking the actual analysis, where we analyze each module separately and then combine the constraints, we simplify the procedure by considering a single spectrum (rather than all four) per mock analysis to speed up the computation. We generate 100 Monte Carlo (MC) spectra with no DM line, using the \textsc{fakeit} tool in \textsc{xspec}. Each spectrum has 200 ks exposure, and is generated using the best-fit spectral model of FPMA, obsID 40410001002. This simplification is motivated by the fact that the spectrum for each module has similar best-fit model parameters, and hence statistics. We also test the results obtained with 10 of these simplified simulations against 10 full realizations (i.e., including both obsIDs and both FPMs) and find good agreement. We then pass these mock spectra through the same fitting and line-search procedure as the data. At each mass bin, we thus have 100 simulated upper limits. We interpolate the cumulative distribution of these upper limits and find the corresponding 68\% and 95\% intervals. The upper limits can then obtained directly from the line-search procedure (see Sec.~\ref{sec:dm_search}) without needing to combine different FPMs.
\par The right panel of Fig.~\ref{fig:dmgamma_both_figs} shows the expected upper limit bands obtained with the mock spectra. Our upper limits obtained from real data are consistent with the MC expectation across most of the 10--40\,keV mass range at the $2\sigma$ level; however, there are several features that warrant closer attention.
\par As described in Sec.~\ref{sec:spect_fit}, the high $\chi^2$ values for spectra 40410001002A and 40410002002B are caused primarily by isolated excesses in the energy ranges ${\sim}\text{15--20\,keV}$ and ${\sim}$8--9\,keV, respectively. We first consider the possibility of these excesses being purely statistical. Though the corresponding $p$-values are small---0.06 and 0.03, respectively---this possibility is supported by these excesses appearing in only two of the spectra, and in two different energy ranges. Additionally, the upward fluctuations in the observed limit lie within the 95\% band expected from MC simulations incorporating only statistical fluctuations (see the right panel of Fig.~\ref{fig:dmgamma_both_figs}). If we consider the extreme procedure of excluding the energy ranges 15--20\,keV and 8--9\,keV in spectra 40410001002A and 40410002002B, respectively, the DM limits in the mass ranges ${\sim}$16--18\,keV and ${\sim}$30--40\,keV are strengthened by a factor ${\sim}1.3$, as we are no longer including data which favor nonzero DM flux; excluding these excesses also reduces the best-fit continuum level over the rest of the energy range, slightly weakening the overall limit elsewhere by at most a factor ${\sim}1.5$. In both cases, the changes in the DM limit are well within the MC band of Fig.~\ref{fig:dmgamma_both_figs}, so we do not pursue the extreme procedure of excluding these energy ranges from our analysis \textit{a posteriori}.
\par We also test the effect of incorporating a flat $7.5\%$ systematic across the entire energy range of all four spectra, sufficient to give $\chi^2/60 \lesssim 1$ for each. We run these spectra through the same line-search procedure as our default analysis, and find a combined DM limit that is a factor ${\sim}1.5$ weaker than our default result, but still a factor ${\sim}1.5$ stronger than the previously-leading Ref.~\cite{Neronov:2016wdd} in the mass range 10--12\,keV. We conclude that any systematic effects on our final DM limit are subdominant to the range expected from statistical fluctuations already shown in the right panel of Fig.~\ref{fig:dmgamma_both_figs}. Adding such a flat systematic to all four spectra is also an extreme procedure considering the two spectra lacking these excesses (40410001002B and 40410002002A), as well as the isolated energy ranges in which these excesses appear; therefore, we do not apply such a flat systematic when calculating our default DM limit.
\par It is plausible that the excesses described previously result from some un-modeled, transient background component. Such a component was not evident during our initial data screening (see Sec.~\ref{sec:newGC}), but there known issues with the default \nustar{} background model in these regions. (As noted previously, the excesses in the 8--9\,keV and 15--20\,keV energy ranges are inconsistent with DM.) If we were to add additional background components in these regions, our DM limit in those regions would become stronger, as some of the flux assigned to the DM line would instead be incorporated into the new background components. Elucidating the form of these additional background components---if they exist---is beyond the scope of this work, and will require analysis (ongoing) of \nustar{} datasets with significantly longer exposure time.
\par We conclude by considering the ranges where the DM limit in Fig.~\ref{fig:dmgamma_both_figs} most departs from the MC expectation, though in all cases the observed limit remains consistent with the 95\% MC band. (The upward fluctuations in the observed limit near masses ${\sim}$16--18\,keV and ${\sim}$30--40\,keV have already been discussed.) First, the upward fluctuations near the edges of the region of interest (masses 10\,keV and 40\,keV) likely arise from parts of the DM line leaving the energy range 5--20\,keV. Second, the upward fluctuation in the MC band near $m_\chi \simeq 20\text{\,keV}$ is attributed to a weak line near $E \simeq 10\text{\,keV}$ in the background model, whereas the observed limit exhibits a downward fluctuation due to negative residuals in spectrum 40410001002B at $E \simeq 10\text{\,keV}$. Finally, we turn to the mass range ${\sim}\text{10--12\,keV}$, where our results improve the most compared to previous analyses and the observed limit also touches the lower end of the MC band. A closer inspection shows that this is driven by several downward-fluctuating data points from 40410001002A and 40410002002A/B. These negative residuals appear at different energies in three different modules, and the bin widths are a factor ${\sim}4$ narrower than the detector energy resolution. This lends support to the strong limit being caused by statistical downward fluctuations. 

\subsection{Sterile-Neutrino DM Constraints}
\label{sec:constraints}
For sterile-neutrino DM, we convert the decay rate constraints to mixing angle constraints using \cite{Shrock:1974nd,Pal:1981rm}
\begin{equation}
	\Gamma = 1.38\times 10^{-32}\,{\rm s^{-1}} \left(\frac{\sin^{2}2\theta}{10^{-10}}\right) \left(\frac{m_{\chi}}{\rm\,keV}\right)^{5}.
\end{equation}
The aggregate constraints in the mass-mixing-angle plane from x-ray searches (including \nustar{}) are shown in Fig.~\ref{fig:SNDM_NuSTAR_best}. As described previously, our high-latitude Galactic bulge constraints are a factor ${\sim}2$ stronger than the previous leading limits \cite{Neronov:2016wdd} in the mass range 10--12\,keV while requiring a factor ${\sim}50$ less exposure time, and are comparable with previous \nustar{} constraints over the rest of the 10--40\,keV mass range. This supports the use of observation regions with low astrophysical background and large J-factors.

In the context of the $\nu$MSM, the parameter space is also bounded by production and structure formation constraits~\cite{Venumadhav:2015pla, Cherry:2017dwu}~(see also Ref.~\cite{Ng:2019a} for discussion). As discussed in Sec.~\ref{sec:monte_carlo}, the DM line analysis in this paper is limited mostly by statistics, except for the known feature near $E\simeq 15\text{\,keV}$. To cover the $\nu\text{MSM}$ window for $m_\chi > 10\text{\,keV}$, a factor ${\sim}4$ improvement in sensitivity is needed, corresponding to ${\sim}4$ Ms exposure of regions with large J-factors and minimal astrophysical backgrounds (similar to the present paper). Though a survey of this depth is feasible, we caution that systematic deviations from the default \nustar{} background model will likely prevent long exposures from reaching their design sensitivity until an improved model of the \nustar{} instrumental background can be developed. Ongoing work for improving the \nustar{} instrumental background model, especially in the 3--5\,keV energy range, will be essential for further testing of the $\nu$MSM down to ${m_\chi = 6\text{\,keV}}$, including the tentative signal at $E\simeq 3.5$\,keV.


\section{\label{sec:conclusions} Conclusions and Outlook}
\par The \nustar{} observatory's large FOV for unfocused x-rays has been pivotal in constraining the properties of sterile-neutrino DM with $m_\chi \sim \text{keV}$, such as that predicted by the $\nu\text{MSM}$. \nustar{} observations of the Galactic center, blank-sky extragalactic fields, and M31 have provided world-leading constraints on the ${\chi \rightarrow \nu+\gamma}$ decay rate in the mass range 10--50\,keV, practically closing the ``window'' in the $\nu\text{MSM}$ parameter space for masses 20--50\,keV. Closing the window for masses $\text{6--20\,keV}$, however, has proved difficult, due to large astrophysical x-ray backgrounds in the observation regions.
\par In this paper, we analyze a combined ${{\sim}190\text{\,ks}}$ of \nustar{} observations to search for x-rays originating from the radiative decay of sterile-neutrino DM in the Galactic halo. The observation regions were optimized to reduce astrophysical x-ray backgrounds from Galactic x-ray sources and from the Galactic ridge x-ray emission while remaining near the center of the Galactic halo, where the DM decay signal is expected to be strongest. We consistently model the flux from both the focused (2-bounce) and unfocused (0-bounce) \nustar{} apertures, though our sensitivity to decaying DM is dominated by the large unfocused FOV. To avoid the systematic effects of stacking spectra with different instrumental and astrophysical backgrounds, we model the spectra individually and combine the sensitivity of each.
\par Finding no evidence of sterile-neutrino DM decays, we instead set upper limits on the sterile neutrino decay rate in the mass range 10--40\,keV. In the mass range ${{\sim}\text{10--12\,keV}}$, our limits are a factor ${\sim}2$ stronger than the previous leading limits while requiring a factor ${\sim}50$ less exposure time. This is due in part to the low astrophysical background and large J-factor in these optimized observation regions, as well as downward statistical fluctuations. We also perform Monte Carlo simulations to determine our expected DM sensitivity, and find that our derived limits are consistent with expectations across most of the 10--40\,keV mass range. 
\par As the astrophysical background (now dominated by the irreducible CXB flux) in these observations is comparable to the instrumental background, we observe deviations of the spectra from the default \nustar{} background model, particularly in the energy ranges ${{\sim}\text{8--9\,keV}}$ and ${{\sim}\text{15--20\,keV}}$. Though similar effects are visible in other \nustar{} analyses (see the left panel of Fig.~\ref{fig:dmgamma_both_figs}), the excesses in our spectra are consistent with statistical fluctuations (see Sec.~\ref{sec:monte_carlo}). Detailed characterization of the instrumental background is ongoing, and additional \nustar{} searches, particularly with an improved model of the instrumental background, will be uniquely suited to probing the remaining $\nu\text{MSM}$ parameter space, as well as investigating the nature of the $\text{3.5-keV}$ line.


\section*{\label{sec:acknowledgements} Acknowledgments}
\par We thank Alexey Boyarsky, Steve Rossland, Oleg Ruchayskiy, and Shuo Zhang for helpful comments and discussions. We also thank the anonymous referees for their constructive comments.
\par The \nustar{} observations described in this work were awarded under NASA Grant No. 80NSSC18K1615. We thank the \nustar{} team at NASA, JPL, and CalTech for the excellent performance of the instrument and their assistance with initial data processing.
\par The computational aspects of this work made extensive use of the following packages: \textsc{saoimage ds9} distributed by the Smithsonian Astrophysical Observatory; the \textsc{scipy} ecosystem \cite{scipy}, particularly \textsc{matplotlib} and \textsc{numpy}; and \textsc{astropy}, a community-developed core \textsc{python} package for Astronomy \citep{astropy:2013, astropy:2018}. This research has made use of data and software provided by the High Energy Astrophysics Science Archive Research Center (HEASARC), which is a service of the Astrophysics Science Division at NASA/GSFC and the High Energy Astrophysics Division of the Smithsonian Astrophysical Observatory.
\par B.M.R. and K.P. receive support from NASA Grant No. 80NSSC18K1615. B.M.R. is also partially supported by MIT Department of Physics and School of Science fellowships. K.C.Y.N.\ is supported by a Croucher Fellowship and a Benoziyo Fellowship. K.P.\ receives additional support from the Alfred P.\ Sloan Foundation and RCSA Cottrell Scholar Award No. 25928. J.F.B.\ is supported by NSF Grant No.\ PHY-1714479. S.H.\ is supported by the U.S.\ Department of Energy under Award No.\ DE-SC0018327, as well as NSF Grants No.\ AST-1908960 and PHY-1914409. R.K.\ receives support from the Russian Science Foundation under Grant No.\ 19-12-00396. D.R.W. is supported by NASA ADAP Grant No. 80NSSC18K0686.

\pagebreak

\bibliography{bib.bib}

\end{document}